\documentclass[12pt]{article}
\usepackage{latexsym}
\let\a=\alpha\let\b=\beta\let\d=\delta
\let\g=\gamma
\let\l=\lambda

\let\s=\sigma\let\t=\tau

\def\ket#1{|#1\rangle}

\def\IR{\relax{\rm I\kern-.18em R}}
\def\IC{\relax\hbox{$\inbar\kern-.3em{\rm C}$}}
\def\IZ{\relax{\rm Z\kern-.5em Z}}

\newcommand{\SL}[0]{{\rm SL}(2,\IR)}

\newcommand{\Lads}{{\rm Lorentzian} \ AdS_3}
\newcommand{\Eads}{{\rm Euclidean} \ AdS_3}

\newcommand{\be}{\begin{equation}}
\newcommand{\ee}{\end{equation}}
\newcommand{\beq}{\begin{equation}}
\newcommand{\eeq}{\end{equation}}
\newcommand{\bea}{\begin{eqnarray}}
\newcommand{\eea}{\end{eqnarray}}

\newcommand{\del}{\partial}

\newcommand{\nbox}{{\,\lower0.9pt\vbox{\hrule \hbox{\vrule height 0.2 cm \hskip
0.2 cm \vrule height 0.2 cm}\hrule}\,}}

\begin{document}
\rightline{hep-th/0109200}
\rightline{UCI-TR 2001-28}
\rightline{RUNHETC-2001-28}
\vskip 1cm
\centerline{\Large \bf New  $AdS_3$ Branes and  Boundary States}
%\centerline{\Large \bf in }
\vskip 1cm

\renewcommand{\thefootnote}{\fnsymbol{footnote}}
\centerline{{\bf \large Arvind
Rajaraman\footnote{arvindra@muon.rutgers.edu}
}}
\vskip .5cm
\centerline{\it Serin Laboratory, Rutgers University, Piscataway, NJ 08854}
%\centerline{\it Piscataway, NJ 08854}
\vskip .2cm
\centerline{and}
\vskip .2cm
\centerline{\it Frederick Reines Hall, University of California, Irvine, CA 92697.}
%\centerline{\it Irvine, CA 92697}

\setcounter{footnote}{0}
\renewcommand{\thefootnote}{\arabic{footnote}}

\begin{abstract}
We examine D-branes on $AdS_3$, and find a three-brane wrapping
the entire $AdS_3$, in addition to 
1-branes and instantonic 2-branes previously 
discussed in the literature. The three-brane
is found using a construction of Maldacena, Moore, and
Seiberg. We show that
all these branes satisfy Cardy's condition and extract the
open string spectrum on them.
\end{abstract}

\section{Introduction}
Perturbative string theory on $AdS_3$, which is isomorphic
to the $\SL$ group manifold, is considerably more
complicated than string theory on compact group manifolds.
(For previous work on $\SL$, see \cite{
Petropoulos:1999me,Petropoulos:1999nc,
noghost,Giveon:1998ns,Gawedzki:1991yu,
deBoer:1998pp,
Teschner:2000ug,
Kato:2000tb}.)
Some of the subtleties of the closed string spectrum
were worked out in \cite{mo}, where a proposal for
the closed string spectrum was proposed, and checked
by an explicit computation of the partition function.

There has also been work on D-branes and
open strings in $AdS_3$. Semiclassical
descriptions of D-branes were given in
\cite{Stanciu:1999nx,
Figueroa-O'Farrill:2000ei,bachas},
and the direct quantization of
the open strings on the branes was performed
in \cite{ooguri}. The boundary state
description of the D-branes was
found in \cite{us}.

In this paper, we will extend the results
of  \cite{us} to describe boundary states for other branes on
$AdS_3$ and $\Eads$. (Semiclassical descriptions
of some of these branes had been given in \cite{Stanciu:1999nx,
Figueroa-O'Farrill:2000ei,bachas}).
We verify that Cardy's condition is
satisfied and use this to extract the spectrum
of open strings connecting various branes.

The interesting aspect in these calculations is
that the infinite volume of $AdS_3$ makes
quantities like the overlaps of branes infinite.
As we shall see (and was already shown in \cite{us})
all these divergences can be tamed by
a careful regularization of the volume.

We start by reviewing the closed string spectrum of
\cite{mo}. Then we review the
D-brane spectrum on $\Eads$, which is 
a straightforward extension of \cite{Stanciu:1999nx,
Figueroa-O'Farrill:2000ei,bachas}.
We find a whole zoo of objects, which fall
into three classes. 
The boundary states for these branes can be
written down, following the methods of
\cite{us}, and the overlaps can be computed.

It turns out that one
of these three classes  actually fails to
satisfy Cardy's condition. The remaining
two classes of objects (which we call 1-branes
and 2-instantons) satisfy Cardy's condition
and hence are physical states in the
theory.

We then investigate a new class of branes 
(which we call {\it parafermionic} branes)which
were not discussed by  \cite{Stanciu:1999nx,
Figueroa-O'Farrill:2000ei,bachas}.
These new branes are the extension to $\SL$ of
branes found in $SU(2)$ by \cite{seib}.
This yields one new brane that satisfies Cardy's
condition and appears to be a three-brane 
wrapping the entire $AdS_3$.

We close with a summary of our results.

Recently, the paper \cite{recent} appeared with related results.

\section{Review of closed strings on $AdS_3$}
In cylindrical (global) coordinates, the metric of $AdS_3$ is

\beq
ds^2 = - \cosh^2\rho\ dt^2 + d\rho^2 + \sinh^2\rho\ d\phi^2
\eeq

All modes of particles propagating in $AdS_3$ should fall into representations
 of the isometry group, namely $\SL_L\times \SL_R$.
Explicit expressions for the generators of
 $\SL_L\times \SL_R$ are \cite{vijay}
\bea
\label{gens}
J^3&=& \frac{i}{2} \partial_u \nonumber \\
J^+ &=&\frac{i}{2} e^{-2iu}\left[ \coth 2\rho \partial_u-
\frac{1}{\sinh 2\rho } \partial_v + i \partial_\rho \right] \nonumber\\
J^-&=&
\frac{i}{2} e^{2iu}\left[ \coth 2\rho \partial_u-
\frac{1}{\sinh 2\rho }\partial_v- i \partial_\rho \right]
\eea
where $u = \frac{1}{2} (t+\phi), v=  \frac{1}{2} (t-\phi)$.
%\beq
%u = \frac{1}{2} (t+\phi)~~~~~~~~~~~~
%v=  \frac{1}{2} (t-\phi)
%\eeq
The generators of the other $\SL$ algebra are obtained by exchanging
$u$ and $v$  in the above expressions.

The Laplacian 
on $\Lads$ is 
given by
\beq
\Box  =
\partial_\rho^2 + 2 \frac{\cosh 2\rho }{\sinh 2 \rho } \partial_\rho
+ \frac{1}{\sinh^2 \rho }\partial_\phi^2 - \frac{1}{\cosh^2 \rho }
\partial_t^2 
\eeq
The Laplacian on $\Eads$ is obtained by replacing $t=i\tau$.

We will consider  (delta function)
normalizable states in $\Eads$. This forces the
eigenvalues of the Laplacian to be of the
form $4j(j-1)$ with $j=\frac{1+i s }{2} $ ($s$ real).
These are the  continuous representations of
the covering group $SL(2,C)$. 

These do not have highest or lowest weight states, and therefore
the spectrum of $J^3$ in this representation
is unbounded from above and from below.
This leads to a divergence in the characters of these
representations.

%\subsection{Spectral flow}
For each
representation of the zero mode algebra $\SL\times \SL$ one can
construct
 a module of the Kac-Moody algebra by a repeated application of oscillator
modes $J^a_{-n}, n>0, a= \pm, 3$.
We denote such representations by $\hat{D}_j^\pm$ and
$\hat{C}_j^\alpha$. These representations were dubbed positive energy representations in \cite{mo}, as the spectrum of $L_0$ is bounded from below.

In \cite{hwang,mo}, a new set of representations of the KM algebra was constructed.
These are obtained from the above representations by an application of the
spectral
flow, defined as :
\beq
J_n^3 \rightarrow J_n^3 - \frac{k}{2}\omega \delta_{n,0} ~~~~
J_n^+ \rightarrow J_{n+\omega}^+ ~~~~
J_n^- \rightarrow J_{n-\omega}^-
\eeq

This preserves the KM algebra, and was conjectured in \cite{mo} to be
a symmetry of the closed string spectrum. The new representations have
$L_0$
unbounded from below, but satisfy a no-ghost theorem, proven in \cite{mo}.
The complete closed string
 spectrum then
 consists of the
representations allowed by the no-ghost theorem above, together with all
their
images under the spectral flow---the so-called flowed representations.

\section{The brane zoo}
\subsection{Classical descriptions}
We now review the D-brane spectrum found in
 \cite{Stanciu:1999nx,
Figueroa-O'Farrill:2000ei,bachas}.
Following \cite{as}, we shall look for D-branes wrapping
regular and twined conjugacy classes. 
%This is
%motivated by work on branes in $SU(2)$ WZW models.

We will use the following quantities, defined in
terms of Euclidean cylindrical coordinates (see the Appendix)
\bea
\label{coorddef}
\sinh\psi\equiv\sinh\rho\sin\phi
\nonumber
\\
\sinh\a\equiv\cosh\rho\sinh\tau
\\
\cosh\l\equiv\cosh\rho\cosh\tau
\nonumber
\eea
\subsubsection{1-branes}
One set of conjugacy classes in $\Lads$ is described
by the equation \cite{bachas}
\bea
\label{ccls1}
\sinh\psi={\rm constant}\equiv \sinh\psi_0
\eea
A brane wrapped on this hypersurface is extended along
the time direction and hence describes a D1-brane
in $\Lads$.

In $\Eads$, the conjugacy class is described by the same equation,
and is a hypersurface in $\Eads$.

In Poincare coordinates on $\Eads$, the corresponding equation
for the conjugacy class is 
\bea
{x\over r}={\rm constant}
\eea

\subsubsection{2-instantons}
A second set of conjugacy classes in $\Lads$ is described
by the equation $\cosh\rho\sin t={\rm constant}$.
A brane wrapped on this hypersurface is
not extended along
the time direction and hence describes an
instantonic D2-brane in $\Lads$.

In $\Eads$, the conjugacy class is described by the
equation 
\bea
\label{ccls2}
\sinh\a={\rm constant}\equiv \sinh\a_0
\eea

In Poincare coordinates in $\Eads$, the corresponding equation
for the conjugacy class is 
\bea
{\tilde{\tau}\over r}={\rm constant}
\eea
This shows the similarity between the 2-instanton and the 1-brane 
above.

Equation (\ref{ccls2}) can be understood as being related to
equation (\ref{ccls1}) by the replacement $\rho\rightarrow
\rho+{i\pi\over2}, (\t+\phi)\rightarrow (\t+\phi),
 (\t-\phi)\rightarrow -(\t-\phi)$, which
leaves $\Eads$ invariant. This is an inner automorphism
of the covering group 
$SL(2,C)$.
%${\rm SL}(2,\IC)$.

\subsubsection{0-instantons}
The  $\Lads$ conjugacy class $\cosh\rho\sin t={\rm constant}$
can be translated to the  conjugacy class $\cosh\rho\cos t={\rm constant}$
by a shift $t\rightarrow t+{\pi\over2}$, which is an inner
automorphism of $\SL$.
This new equation has a different continuation to
$\Eads$, 
i.e.
\bea
\cosh\l={\rm constant}\equiv \cosh\l_0
\eea

A brane wrapped on this hypersurface is also instantonic.
In particular, the case $\l_0=0$
describes a pointlike object at $\rho=0,\tau=0$. 
(This
is, as we shall see, {\it not} the usual D-instanton.)
The more general case $\l_0 >0$
describes a hypersurface in $\Eads$ surrounding
the point $\rho=0,\tau=0$. 

As we shall see, these
branes do not satisfy Cardy's condition, and therefore
do not correspond to physical objects.

These conjugacy classes are related to the ones in the
previous subsection by the shift $\tau\rightarrow \tau
-{i\pi\over2}$, which is an inner automorphism
of the covering group 
$SL(2,C)$.
%${\rm SL}(2,\IC)$.

\subsubsection{Circular branes}
More generally, the $\Lads$ conjugacy class 
\bea
\cosh\rho
\ (\sin t \cos t_0 + \cos t \sin t_0)={\rm constant}
\eea
continues in $\Eads$ to
$\cosh\rho(i\sinh \tau \cos t_0 + \cosh \tau  \sin t_0)={\rm constant}$
which is actually two equations
\bea
\cosh\rho\sinh \tau =C_1\quad\qquad \cosh\rho\cosh \tau=C_2
\eea
This describes a circle centered at
$\rho=0$, at some fixed value of $\tau$.

As we shall see, none of these conjugacy classes lead
to boundary states satisfying Cardy's condition unless
%either 
$\sin t_0=0$, 
%or $ \cos t_0=0$, 
which reduces to the previous
cases.

\subsection{Boundary states}

\subsubsection{1-branes}
The boundary state for the 1-brane was found in \cite{us}. We review this construction
here.

Firstly, the boundary state $\ket{B^{1D}}$ for the 1-brane
satisfies $(J^a+\bar{J}^a)\ket{B^{1D}}=0$.
For example,
the zero-mode equation  $(J^3+\bar{J}^3)\ket{B^{1D}}=0$
translates to $\del_t\ket{B^{1D}}=0$ i.e. the brane is extended along
the time direction. The other components work similarly,
as shown in \cite{us}.

The solution for the Ishibashi states
satisfying  
\beq
\label{cond}
(J^a_n + \bar{J}^a_{-n}) |I \rangle =0
\eeq
was found in \cite{ish}.
Suppose we are given a KM
primary $|\Phi_j\rangle$ in the $\hat{C}_j^{\alpha}
\times \hat{C}_j^{\alpha}$ representation. By that we mean a state
which is annihilated by all lowering operators $J_n^a, n>0$, but not
by   the zero modes (since
$C_j^\alpha$ has no highest weight states).
 Define\cite{contra,armr2}
\be
\label{contra}
|I^j\rangle = \sum_{I,J} M_{IJ}^{-1}J_{-I}
\bar{J}_{-J}|\Phi^j\rangle
\ee
Here $I,J$ are ordered strings of indices $(n_1, a_1)\cdots (n_r,a_r)$, and
\be
J_I=J_{n_1}^{a_1}\cdots J_{n_r}^{a_r}.
\ee

For later convenience we choose an ordering such that all zero modes
act from the left.
This sums over all the descendants in the KM module, with the normalization
 defined as
\be
M_{IJ}=\langle \Phi^j| J_I\, J_{-J}|\Phi^j\rangle
\ee

$M_{IJ}$ is invertible for any KM module. (For degenerate
modules, one has to mod out by the null
vectors).
It is easy to see that $|I^j\rangle$ satisfies (\ref{cond})  by
showing
that $(J_n+\bar{J}_{-n})|I^j\rangle$  is orthogonal to
all states in the module based on
$|\Phi^j\rangle$.

It turns out that it is necessary to separate out the oscillator modes
from the zero-modes (which we sometimes refer to as primaries). This
is because there are divergences in the brane overlaps, which
come from the infinite sum over primaries. We therefore
want to separate out the finite oscillator contribution
from this divergence.

Hence we define a modified Ishibashi state, which is a coherent state
defined as
\beq
\label{ishdef}
|I^j_{m\bar{m}}\rangle_ = \widetilde{\sum_{I,J}} M_{IJ}^{-1}J_{-I}
\bar{J}_{-J}|\Phi^j_{m\bar{m}}\rangle
\eeq

Here $|\Phi^j_{m\bar{m}}\rangle$ is annihilated by all oscillator modes except
for the zero modes (as above), and has magnetic quantum numbers $m$
and $\bar{m}$. The sum
$\widetilde{\sum}$ over descendants here is defined to
exclude any action by the
zero
 modes $J^\pm_0$. It is clear that formally
\beq
\label{ish}
|I^j \rangle = \sum_{m} |I^j_{m,-m}\rangle\
\eeq
 
The Cardy state was found in the large $k$ limit by
requiring the branes to be localized on the conjugacy class in this
limit. 
The boundary state for the D1-brane was then found to be \cite{us}
\bea
\label{1Bdef}
\ket{B^{1D}}=T \sum_s e^{-is\psi_0}\ket{I^s}\equiv T\int{ds\over 2\pi}
e^{-is\psi_0}\ket{I^s}
\eea
where $\ket{I^s}$ is the coherent state defined  in (\ref{ishdef})
\beq
\ket{I^s} = \widetilde{\sum_{I,J}} M_{IJ}^{-1}J_{-I}
\bar{J}_{-J}|\Phi^s\rangle
\eeq
and
$|\Phi^s\rangle$ is the primary satisfying 
$\langle x |\Phi^s \rangle = \frac{e^{is\psi}}{\cosh\psi}$.
($\psi$ is defined in (\ref{coorddef}).)
$T$ is an overall normalization constant. In
the flat space case it was determined by
\cite{DiVecchia:1997pr}.

\subsubsection{2-instantons}
The 
conjugacy class
$\cosh\rho\sinh\t={\rm constant}$ was obtained from the
conjugacy class (\ref{ccls1}) by the continuation
 $\rho\rightarrow
\rho+{i\pi\over2}, (\t+\phi)\rightarrow (\t+\phi),
 (\t-\phi)\rightarrow -(\t-\phi)$. 

This continuation affects the currents as
\bea
J^a\rightarrow J^a, \quad \bar{J}^3\rightarrow -\bar{J}^3,
\quad \bar{J}^+\rightarrow -\bar{J}^-, \quad\bar{J}^-\rightarrow -\bar{J}^+
\eea

Accordingly the Ishibashi state (\ref{ishdef}) is replaced by
(cf. \cite{Brunner:1999fj})
\beq
\label{ish2def}
|\widehat{I^j_{m\bar{m}}}\rangle_ = \widetilde{\sum_{I,J}} M_{IJ}^{-1}K_{-I}
\bar{K}_{-J}|\Phi^j_{m\bar{m}}\rangle
\eeq
with
\bea
K^a_n=J^a_n,\quad \bar{K}^3_n= -\bar{J}^3_n,\quad
\bar{K}^+_n= -\bar{J}^-_n,\quad \bar{K}^-_n= -\bar{J}^+_n
\eea

The boundary state for the D2-instanton can be
written in analogy to  (\ref{1Bdef})
\bea
\label{2Idef}
\ket{B^{2I}}=T \sum_p e^{-ip\a_0}\ket{\widehat{I^p}}\equiv T \int{dp\over 2\pi}
e^{-ip\a_0}\ket{\widehat{I^p}}
\eea
where $\ket{\widehat{I^p}}$ is the modified Ishibashi state (\ref{ish2def}) built over the
primary satisfying 
$\langle x |\Phi^p \rangle = \frac{e^{ip\alpha}}{\cosh\alpha}$.
($\a$ is defined in (\ref{coorddef}).)

We now consider the flowed representations. It was argued in \cite{us} that the
1-brane did not couple to flowed representations. 
Heuristically, the flowed representations represent
strings winding near the boundary of $AdS_3$. The 1-brane 
does not wind near the boundary and hence cannot couple to
these states. It might seem that the 2-instanton can couple
to these flowed representations. However, the symmetry
between the 1-brane and the  2-instanton in Poincare coordinates
shows that in fact there is no such coupling.

 This result
is also a consequence of Cardy's condition. The
overlap of the 2-instantons should be the
same as that of the 1-branes (again due
to the symmetry in Poincare coordinates.) If
the 2-instanton couples to flowed representations, this
matching is destroyed. Hence
we conclude that the  2-instanton does
not couple to flowed representations.

%The full  boundary state 
%is therefore
%can be
%written
%\bea
%\label{2Idef}
%\ket{B^{2I}}=T \sum_{w\e \IZ} \int{dp\over 2\pi}
%e^{-ip\a_0}\ket{\widehat{I^{p,w}}}
%\eea
%where $\ket{\widehat{I^{p,w}}}$ is the modified Ishibashi state (\ref{ish2def}) built over the
%primary $\ket{\Phi^{p,w}}$ which is the image of
%$\ket{\Phi^{p}}$ under spectral flow.

\subsubsection{0-instantons}
The
conjugacy class
$\cosh\rho\cosh \t={\rm constant}$ was obtained from the
conjugacy class $\cosh\rho\sinh \t={\rm constant}$ by the further 
shift
$\tau\rightarrow \tau-{i\pi\over2}$. This shifts the
currents as
\bea
J^3\rightarrow J^3, \quad J^+\rightarrow iJ^+,\quad J^-\rightarrow -iJ^-
\nonumber
\\
\bar{J}^3\rightarrow \bar{J}^3,\quad \bar{J}^+\rightarrow i\bar{J}^+,
\quad \bar{J}^-\rightarrow -i\bar{J}^-
\eea
The Ishibashi state corresponding to a primary $\ket{\Phi^j_{m\bar{m}}}$
is therefore
\beq
\label{ish3def}
|\widehat{\widehat{I^j_{m\bar{m}}}}\rangle_ = \widetilde{\sum_{I,J}} M_{IJ}^{-1}K_{-I}
\bar{K}_{-J}|\Phi^j_{m\bar{m}}\rangle
\eeq
with
\bea
K^3_n=J^3_n, \quad
K^+_n=iJ^+_n, \quad
K^-_n=-iJ^-_n, 
\nonumber
\\
\bar{K}^3_n= -\bar{J}^3_n,\quad
\bar{K}^+_n= i\bar{J}^-_n, \quad
\bar{K}^-_n= -i\bar{J}^+_n
\eea

The boundary state for the D0-instanton can then be
written as
\bea
\ket{B^{0I}}=T \sum_qe^{-iq\l_0}\ket{\widehat{\widehat{I^q}}}\equiv T \int{dq\over 2\pi}
e^{-iq\l_0}\ket{\widehat{\widehat{I^q}}}
\eea
where $\ket{\widehat{\widehat{I^q}}}$ is the modified Ishibashi state (\ref{ish3def}) built over the
primary satisfying
$\langle x |\Phi^q \rangle = \frac{e^{iq\l}}{\sinh\l}$.
($\l$ is defined in (\ref{coorddef}).)

%On including spectral flow, the full  boundary state can be
%written
%\bea
%\ket{B^{0I}}=T \sum_{w\e \IZ}
%\int{dq\over 2\pi}
%e^{-iq\l_0}\ket{\widehat{\widehat{I^{q,w}}}}
%\eea
%where $\ket{\widehat{\widehat{I^{q,w}}}}$ is the 
%modified Ishibashi state (\ref{ish3def}) built over the
%primary $\ket{\Phi^{q,w}}$ which is the image of
%$\ket{\Phi^{q}}$ under spectral flow.

\subsubsection{Circular branes} 
It does not seem possible to write a boundary state for
these branes using the method of \cite{us}. They do not have a good
flat space limit either. It is
likely that these conjugacy classes do not correspond
to physical branes.

\section{Overlaps of branes}
\subsection{General remarks}

The overlap of the branes $\ket{B}$ and
$\ket{\tilde{B}}$ is defined as 
$\langle\tilde{B}|q^{L_0+\bar{L}_0-{c\over 12}}\ket{B}$.

We shall find it useful to separate the oscillator contribution
from the
contribution of the primaries. To do this we define
new coherent states which only have overlap with the primaries.
\bea
\ket{C^{1D}}=\sum_s e^{-is\psi_0}\ket{\Phi^s}
\nonumber
\\
\ket{C^{2I}}=\sum_p e^{-ip\a_0}\ket{\Phi^p}
\\
\ket{C^{0I}}=\sum_qe^{-iq\l_0}\ket{\Phi^q}
\nonumber
\eea

Note that
\bea
\langle x\ket{C^{1D}}=\delta(\sinh\psi-\sinh\psi_0)
\nonumber
\\
\langle x\ket{C^{2I}}=\delta(\sinh\alpha-\sinh\a_0)
\\
\langle x\ket{C^{0I}}=\delta(\cosh\l-\cosh\l_0)
\nonumber
\eea
which shows that the branes are localized on conjugacy classes.

We will often use the relation
\bea
\label{intform}
\langle\tilde{C}|q^{L_0+\bar{L}_0-{c\over 12}}\ket{C}
=\int dx\ \langle\tilde{C}\ket{x} \langle x |q^{L_0+\bar{L}_0-{c\over 12}}\ket{C}
\eea

We will also need the formula:
\beq
\left(L_0 + \bar{L}_0 -\frac{c}{12}\right)  |\Phi^s \rangle
= \left(\frac{-2j(j-1)}{k-2} - \frac{k}{4(k-2)}\right)  |\Phi^s \rangle =
\left(\frac{s^2}{2(k-2)} -\frac{1}{4}\right) |\Phi^s \rangle
\eeq

%Also we define $q=e^{i\pi \tau}$.

\subsection{Preliminary comments on 0-instantons}

We now present our first evidence that the 0-instantons
are not physical states. 

We start with the case $\l_0=0$, which describes a pointlike object, which
at first glance may appear to be the standard D-instanton.
This is in fact not the case.

To see this, note that the usual D-instanton looks like
a ten dimensional delta function. The brane wrapped on the
conjugacy class instead
satisfies
\bea
\langle x\ket{C^{0I}}=\delta(\cosh\l-1)
\eea
and hence
\bea
\int dx \langle x\ket{C^{0I}} =\int d\l \sinh^2\l \ \delta(\cosh\l-1)
=0
\eea
Hence the normalization is incorrect for this state to be a D-instanton.
Similarly the other 0-instantons do not resemble
known D-branes in the flat space limit.

As we shall now see, the 0-instantons do not satisfy Cardy's condition either.
To see this, we shall compute the overlap of
a 2-instanton with a 0-instanton.

\subsection{Overlaps of 2-instantons and 0-instantons
}
The 0-instanton will be labelled by $\l_0$, and
the 2-instanton will be labelled by $\a_0$.

The overlap of primaries (using (\ref{intform})) is
\bea
\langle\tilde{C^{0I}}|q^{L_0+\bar{L}_0-{c\over 12}}\ket{C^{2I}}
%~~~~~~~~~~~~~~~~~~~~~~~~~~~~~~~~~~~~~~~~~~~~~~~~~~~~~~~~~~~~~~~
%~~~~~~~~~~~~~~~~~~~~
%\nonumber
%\\
=\int_0^\infty d\l \int_{-\pi/2}^{\pi/2}  d\s\int_0^{2\pi}d\phi
\sinh^2\l \cos\s
%\int {ds\over 2\pi}
%\d(\cosh\l-\cosh\l_0)  
%~~~~~~~~~~~~~~~~~~~~~~~~~~~~~~~~
~~~~~~~~~~~~~~~~~~~~~~~~~~~
\nonumber
\\
~~~~~~~~~~~~~~~~~~~~~~~~~~~~~~~~\times \d(\cosh\l-\cosh\l_0)  
\int {ds\over 2\pi}
 q^{-\frac{1}{4}}
q^{\frac{s^2}{2(k-2)}}
%e^{-i\tilde{\a}_0s}
{e^{i({\a}-{\a}_0)s}\over \cosh\a}
\eea

For $\l_0=0$, the integral is easily seen to be zero.

More generally, changing variables from $\s$ to $\a$, the
integral is seen to be
\bea
\langle\tilde{C^{0I}}|q^{L_0+\bar{L}_0-{c\over 12}}\ket{C^{2I}}
=2\pi q^{-\frac{1}{4}} \int_{-\l_0}^{\l_0} d\a \sqrt{\frac{k-2}{2i\pi^2\tau}}
\exp\left({ (\a-
\a_0)^2 (k-2)\over  2i\pi \tau}\right)
\eea
%To compute the oscillator contribution, we note that for
%any action of the oscillators $J^3_{-n}, J^{\pm}_{-n}$ with
%$n>0$, there is precisely one combination of
%$J^{\pm}_{0}$ which allows the resulting state to have 
%zero eigenvalue under $J^3-\bar{J}^3$. All these
%states contribute with eual weight to the Ishibashi state.

The oscillators contribute an
extra factor
$[\prod_{n=1}^\infty (1-q^{2n})]^{-3}$.

%(EXTRA FACTORS?)

Combining this with the contribution from the primaries,
we find the overlap of
the boundary states. This resulting expression does
not yield a good open string partition function upon
modular transformation. 
%Including spectral flow
%does not
%change this result.

This indicates that both the 2-instanton and the 0-instanton cannot be 
included consistently in the theory. Since the
2-instanton is related to the 1-brane by flipping $x,\tilde{\tau}$ in
Poincare coordinates, we should have
both these branes in the theory. On the other hand,
we have seen that the 0-instanton does not have a good
flat space limit. Hence we find that the 0-instantons should be
discarded.

\subsection{Overlaps of 2-instantons and 1-branes}
\label{subsec1}

We now turn to the overlap of a 2-instanton and a 1-brane.
We need to regulate a divergence, because
the intersection of the D1-brane and the 2-instanton is the common solution in
$AdS_3$ of the equations $\cosh\rho\sinh\tau={\rm const}$
and $\sinh\rho\sin\phi={\rm const}$, which has an infinite volume. Indeed,
the volume of the intersection region is
\bea
\int dx \ \d(\sinh\a-\sinh\a_0) \ \d(\sinh\psi-\sinh\psi_0)
~~~~~~~~~~~~~~~~~~~~~~~~~~~~~~~~~~~~~~
\nonumber
\\
=\int {d\rho d\psi d\a\  \sinh\rho\cosh\rho\cosh\psi\cosh\a\over 
\sqrt{(\sinh^2\rho-\sinh^2\psi)(\cosh^2\rho+\sinh^2\a)}}
~~~~~~~~~~~~~~~~~~~~
\nonumber
\\
~~~~~~~~~~~~~~~~~~~~~~~~~~~~~~~~~~~~~~~~~~~~~
 \times \d(\sinh\a-\sinh\a_0)\ \d(\sinh\psi-\sinh\psi_0)
\eea
For large $\rho$ (at fixed $\a,\psi$), there is a volume divergence as the
volume grows as
\bea
V\sim \int^{\rho_0} d\rho \sim \rho_0
\eea
This implies that the open string partition function will
have a volume divergence. It can be written
as
\bea
Z_{open}\sim \rho_0 \tilde{Z}
\eea
where $\tilde{Z}$ is finite.

Now since the open string partition function
is divergent, for consistency the overlap of
the branes on the closed string side should
also be divergent with a leading divergence
proportional to $\rho_0$, and as we shall see,
this is true. To check Cardy's condition, we
can drop the common divergent factor  $\rho_0$, and
examine the modular transformation of
the remaining piece. 
%On the closed string side, the overlap of the branes is
%also divergent. To compare the finite quantities, we need to extract the
%leading divergence in the overlap.

The overlap of primaries (using (\ref{intform})) is
\bea
\langle\tilde{C^{2I}}|q^{L_0+\bar{L}_0-{c\over 12}}\ket{C^{1D}}
=
\int {d\rho d\psi d\a\ \sinh\rho\cosh\rho\cosh\psi\cosh\a\over
\sqrt{(\sinh^2\rho-\sinh^2\psi)(\cosh^2\rho+\sinh^2\a)}}
~~~~~~~~~~~~~~~~~~~
\nonumber
\\
%~~~~~~~~~~~~~~~~~~~~~~
\times \  \d(\sinh\a-\sinh\a_0)
\int {ds\over 2\pi} q^{-\frac{1}{4}} q^{\frac{s^2}{2(k-2)}}
%e^{-i\tilde{\psi}_0s}{e^{i\tilde{\psi}s}
{e^{i(\psi-{\psi}_0)s}\over \cosh\psi}
\eea
For large $\rho$ at fixed $\a,\psi$, this behaves as
\bea
\int^{\rho_0} {d\rho d\psi d\a}\ \d(\a-\a_0)
\int {ds\over 2\pi} q^{-\frac{1}{4}} q^{(\frac{s^2}{2(k-2)})}
e^{i(\psi-\tilde{\psi}_0)s}
\sim q^{-\frac{1}{4}}\rho_0
\eea
The oscillator contribution comes from states with $J^3=\bar{J}^3=0$,
since only they have overlap with both the 1-brane and the
2-instanton. They occur with a relative phase in the two
Ishibashi states; for every action of $\bar{J}^3_n$ or  $\bar{J}^2_n$
$(n\ge 0)$,
the relative phase is multiplied by $(-1)$.
The oscillators therefore contribute a factor
\bea
{1\over \prod_{n=1}^\infty (1-q^{2n})(1+q^{2n})(1+q^{2n})}
\eea

After modular transformation, we find
that
\bea
\tilde{Z}\propto{1\over \sqrt{ln(q)}\prod_{n=1}^\infty (1-q^{2n})(1-q^{2n-1})(1-q^{2n-1})}
\eea
which we interpret as being the partition function of three bosons, one
with Neumann boundary conditions, and the other two satisfying
Neumann-Dirichlet boundary conditions.
(There is no residual current algebra, since the boundary conditions
of the 1-brane and the 2-instanton are incompatible.)

\section{Parafermionic branes}

We start by reviewing the $SU(2)$ case. Here the A-branes were
constructed by \cite{cardy}. They can be written
\bea
\ket{A,\hat{j}}_C=
\sum_j {S^j_j\over \sqrt{S^j_0}}\ket{A,j}\rangle
\eea
Here the LHS is the Cardy state, while
$\ket{A,j}\rangle$ is the $SU(2)$ Ishibashi state.

The idea of \cite{seib} is to use a different CFT realization of the $SU(2)$ theory using
parafermions. The Hilbert space of the $SU(2)$ WZW model can be decomposed
as $SU(2)_k=(U(1)_k\times (PF)_k)/\IZ_k$, where $(PF)_k$ is the
parafermion theory.

There is a corresponding decomposition of the Cardy states
into Cardy states of the parafermion theory and the $U(1)$ theory
\cite{seib}
\bea
%\ket{A,j}\rangle=\sum_{n=1}^{2k} {1+(-1)^{2j+n}\over 2}
%\ket{A,j,n}\rangle_{PF}\ket{A,n}\rangle_{U(1)}
%\\
\ket{A,\hat{j}}_C={1\over \sqrt{k}}\sum_{\hat{n}}
\ket{A\hat{j}\hat{n}}^{PF}_C \ket{A\hat{n}}^{U(1)}_C
\eea
The detailed description of the Ishibashi states and Cardy states can be
found in \cite{seib}. We will not need the detailed knowledge of
the parafermionic theory.

We can now perform a T-duality on the $U(1)$ theory. 
This exchanges Dirichlet and Neumann conditions. The A-brane
above is the one satisfying Dirichlet conditions
\bea
(\del_\zeta-\del_{\bar{\zeta}})\ket{A\hat{n}}^{U(1)}_C=0
\eea
(we are representing the $U(1)$ theory by a free boson $\zeta$.)
%They can be explicitly represented as
%\bea
%\ket{A,n}\rangle_{U(1)}=\exp\left( \sum_{n=1}^{\infty}{1\over n}
%\a_{-n}\tilde{\a}_{-n}\right) \sum_{l} \ket{
%{r+2kl\over \sqrt{2k}},{r+2kl\over \sqrt{2k}}}
%\\
%\ket{A\hat{n}}^{U(1)}_C={1\over (2k)^{1/4}}\sum_{n=0}^{2k-1}
%e^{-\pi \hat{n} n\over k}\ket{A,n}
%\eea
%and $\ket{A,n}$ is the Ishibashi state based on the primary
%labelled by $n$.
%These are interpreted geometrically as D0 branes sitting
%at $2k$ special points on the circle.

The branes with Neumann boundary conditions are the B-branes
satisfying
\bea
(\del_\zeta+\del_{\bar{\zeta}})\ket{B,\eta}^{U(1)}_C=0
\eea
%These are D-branes wrapping the circle, with special values of
%the Wilson line.

Thus, after T-duality, we find a new set of branes
in the $SU(2)$ theory \cite{seib}
\bea
\ket{B,\hat{j},\eta=\pm1}_C={1\over\sqrt{k}}\ket{B,\eta}^{U(1)}_C
\sum_{\hat{n}=0}^{2k-1}\ket{A\hat{j}\hat{n}}^{PF}_C
\eea

\subsection{Shape of the branes}
The shape of these branes can be found in the semiclassical limit.

The primaries of the $SU(2)$ theory can be written
\bea
\Phi^j_{m\bar{m}}=V^j_{m\bar{m}}e^{im\zeta+i\bar{m}\bar{\zeta}}
\eea
where $V^j_{m\bar{m}}$ is an operator in the parafermion theory
and $\zeta$ is a free boson.
 
In  coordinates where the metric of $SU(2)$ is
$ds^2=d\eta^2+\cos^2\eta d\g^2+\sin^2\eta d\theta^2$
$m+\bar{m}$ and $m-\bar{m}$ are the momenta conjugate to
translations in $\g$ and $\theta$ respectively.

The B-branes are Neumann with respect to $\zeta$ i.e. they
only couple to states with $m=\bar{m}=0$.
The coupling to these states is the same for both A- and B-branes.

In other words, if the A-brane satisfies in the large $k$ limit
\bea
\langle x\ket{A}=\sum c^j_{m\bar{m}}\langle x\ket{\Phi^j_{m\bar{m}}}
\eea
then the B-brane satisfies
\bea
\langle x\ket{B}=\sum c^j_{0\bar{0}}\langle x\ket{\Phi^j_{0\bar{0}}}
\eea

Hence (upto an overall constant)
\bea
\label{AtoB}
\langle x\ket{B}=\int {d\theta\over 2\pi} {d\g\over 2\pi}\ \langle x\ket{A}
\eea

In  coordinates where the metric of $SU(2)$ is $
ds^2=d\tilde{\psi}^2+\sin^2\tilde{\psi} (d\tilde{\chi}^2+\sin^2\tilde{\chi}
 d\tilde{\omega}^2)$, the 
explicit expression for the
shape of the A-brane is
\bea
\langle x\ket{A} \sim T \ \d(\cos\tilde{\psi}-\cos\tilde{\psi}_0)
\eea

These coordinates are related to the other coordinates through
the relation
\bea
\cos\tilde{\psi}=\cos\eta\sin\g\qquad \sin\tilde{\psi}\sin\tilde{\chi}=\sin\eta\quad
\tilde{\omega}=\theta
\eea

Hence
\bea
\langle x\ket{B}=\int {d\g\over 2\pi}\ T \ \d(\cos\tilde{\psi}-\cos\tilde{\psi}_0)
\nonumber
\\
={T\over \pi\sqrt{2}}{\Theta(\cos\eta-\cos\tilde{\psi}_0)\over \sqrt{(\cos2\eta-
\cos2\tilde{\psi}_0)}}
\eea
in agreement with \cite{seib}.

\subsection{Parafermionic branes in $AdS_3$}
We expect a similar set of new branes in $AdS_3$.
We shall use the intuition from the $SU(2)$ model
to find the shape of these new branes. This
is not guaranteed to work, as the structure
of the primaries in the $\SL$ model is so different.
As we shall see, the naive calculation we do
here does not give the correct answer, but provides a 
guide to it.

%The shape of these branes can be found by the same methods 
%we have already used.
The shape of the A-branes in the large $k$ limit is
\bea
\langle x\ket{A} \sim T \d(\sinh\psi-\sinh\psi_0)
\eea
$\psi$ is related to cylindrical coordinates
through the relation $\sinh\psi=\sinh\rho\sin\phi$.
If we use the formula (\ref{AtoB}), then we find 
the shape of the B-branes to be
\bea
\label{propB}
\langle x\ket{B}\sim T\int {d\phi\over 2\pi} \d(\sinh\psi-\sinh\psi_0)
\nonumber
\\
={T\over \pi}{\Theta(\sinh\rho-\sinh\psi_0)\over \sqrt{\sinh^2\rho-
\sinh^2\psi_0)}}
\eea

These branes can be visualized as covering the whole of $AdS_3$
except for a cylindrical hole in the middle. 

\subsection{Overlaps of the branes}
To check these calculations, we now ask if the branes
satisfy Cardy's condition. First we will look at
the overlap of the A-brane with the B-brane.

The overlaps of the A- and B-branes in $AdS_3$ is divergent and
can be understood in the same way as in section (\ref{subsec1}).

The intersection region has a volume
\bea
\int d\rho d\psi  d\t 
\left({\cosh\rho\sinh\rho\cosh\psi\over \sqrt{\sinh^2\rho-\sinh^2\psi)}}
\right)
\left(T \d(\sinh\psi-\sinh\psi_0) \right)
\nonumber
\\
\times
\left(
{T\over \pi}
{\Theta(\sinh\rho-\sinh\psi_0)\over \sqrt{\sinh^2\rho-\sinh^2\psi_0)}}
\right)
\eea
which is divergent as $\rho$ tends to infinity at fixed $\psi,\t$.
In this region the volume is
\bea
V\sim \int^{\rho_0} d\rho d\psi  d\t  {T^2\over \pi}\d(\psi-\psi_0)
\sim \int d\tau\rho_0  {T^2\over \pi}
\eea

The overlap of primaries is now
\bea
\int d\rho d\psi  d\t
\left({\cosh\rho\sinh\rho\cosh\psi\over \sqrt{\sinh^2\rho-\sinh^2\psi)}}
\right)
\left(
{T\over \pi}
{\Theta(\sinh\rho-\sinh\psi_0)\over \sqrt{\sinh^2\rho-\sinh^2\psi_0)}}
\right)
\nonumber
\\
\times\ T
\int {ds\over 2\pi} q^{-\frac{1}{4}} q^{\frac{s^2}{2(k-2)}}
e^{-i\tilde{\psi}_0s}{e^{i\tilde{\psi}s}\over \cosh\psi}
\eea
As $\rho$ tends to infinity at fixed $\psi,\t$, this tends
to
\bea
\int^{\rho_0} d\rho d\psi  d\t
\int {ds\over 2\pi}
{T^2\over \pi} q^{-\frac{1}{4}} q^{\frac{s^2}{2(k-2)}}
e^{-i\tilde{\psi}_0s}e^{i\tilde{\psi}s}
%~~~~~~~~~~~~~~~~~~~~~~~~~
\nonumber
\\
=\int d\tau {T^2\over \pi} q^{-\frac{1}{4}}\rho_0
\eea

The oscillator contribution comes from $\zeta$, which
has Neumann boundary conditions on the B-brane and
Dirichlet boundary conditions on the A-brane, and
from the parafermions, which have the same boundary
conditions on both branes. Accordingly the
oscillators contribute a factor
\bea
{1\over \prod_{n=1}^\infty (1-q^{2n})(1-q^{2n})(1+q^{2n})}
\eea

After modular transformation, we find that the 
resulting open string partition function can be
understood as the partition function of
three bosons, one having Neumann-Dirichlet conditions,
and the other two satisfying  Neumann  boundary conditions.

Hence the overlap so far is consistent with Cardy's condition.
However, this only tests the proposal (\ref{propB}) in the limit $\rho\sim
\infty$. So we have established so far that
\bea
\label{limB}
\lim_{\rho\to\infty}\langle x\ket{B}=e^{-\rho}
\eea
We now turn to the overlap of the B-brane with itself.
As usual there is a divergence in the overlap coming
from the region $\rho\sim\infty$.

At $\rho\sim\infty$, $e^{-\rho}$ is an
eigenvalue of the Laplacian
\bea
\Box e^{-\rho} ={1\over \cosh\rho\sinh\rho}\del_\rho \left( \cosh\rho\sinh\rho
\del_\rho e^{-\rho}\right)\sim -e^{-\rho}
\eea
with eigenvalue $-1-s^2=-1$. Hence in this limit
\bea
\langle x |q^{L_0+\bar{L}_0-{c\over 12}}\ket{B}\sim q^{-\frac{1}{4}}e^{-\rho}
\eea
In other words, in this limit, only the
primary with $s=0$ contributes.

The overlap of primaries is then
\bea
\int d\rho d\phi  d\t \cosh\rho\sinh\rho\ \langle B\ket{x}
\langle x| q^{L_0+\bar{L}_0-{c\over 12}}\ket{B}
\sim
\int d\t q^{-\frac{1}{4}}\rho_0
\eea

The oscillators contribute the usual factor
$[\prod_{n=1}^\infty (1-q^{2n})]^{-3}$.

The open string partition function is obtained after 
a modular transformation, and can easily be seen to have
the interpretation of the partition function of three free bosons with
Neumann boundary conditions.

This is so far only the leading divergent piece. If
this is to be the exact final answer, then the overlap
of the branes should only come from the $s=0$ state.
In the large $k$ limit, this is the primary
satisfying 
\bea
\Box \Phi=-\Phi
\eea
which is solved by $\Phi={1\over \cosh\rho}\ _2F_1(\frac{1}{2},\frac{1}{2},
1, \tanh^2\rho)$.

In this limit the shape of the B-brane
is therefore
\bea
\langle x\ket{B}\propto {1\over \cosh\rho}\ _2F_1(\frac{1}{2},\frac{1}{2},
1, \tanh^2\rho)
\eea
which is consistent with the limit (\ref{limB}).

We can therefore write the boundary state for
the parafermionic brane
as
\bea
\label{3Bdef}
\ket{B}=T \ \ket{I^0}
\eea
where $\ket{I^0}$ is the coherent state defined  in (\ref{ishdef})
\beq
\ket{I^0} = \widetilde{\sum_{I,J}} M_{IJ}^{-1}K_{-I}
\bar{K}_{-J}|\Phi^0\rangle
\eeq
where $K^1,K^2$ are the parafermionic modes, and
$K^3$ is the free boson $\zeta$.
$|\Phi^0\rangle$ is the primary satisfying
 %$(L_0+\bar{L}_0) |\Phi^0 \rangle = - |\Phi^0 \rangle$.
%\beq
$
\left(L_0 + \bar{L}_0 -\frac{c}{12}\right)  |\Phi^0 \rangle
=\left(-\frac{1}{4}\right) |\Phi^0 \rangle
$
%\eeq

This describes a 3-brane covering the entire
$AdS_3$ space.

In the flat space limit, the brane couples only to modes with
$s^2\sim p^2=0$. Hence it appears locally like a flat space
three-brane.
The other branes which do not cover the whole 
$AdS_3$ do not satisfy Cardy's condition. This
is in agreement with
the flat space limit, where we do not have boundary
states for open branes.

The total integral of the three-form field strength through
the brane is nonzero. However,
the brane is noncompact, so there is no obvious problem.

\section{Summary}

We have discussed several different branes in $AdS_3$. The
ones that satisfy Cardy's condition are
\begin{itemize}
\item{The 1-brane}:\ This is a brane wrapping the conjugacy
class $\sinh\psi={\rm constant}$. The
boundary state for this brane is given in (\ref{1Bdef}).

\item{The 2-instanton}:\ This is a brane wrapping the conjugacy
class $\cosh\rho\sinh\tau={\rm constant}$. The
boundary state for this brane is given in (\ref{2Idef}).

\item{The parafermionic 3-brane}:\  This is a brane wrapping the
whole $AdS_3$ space. The
boundary state for this brane is given in (\ref{3Bdef}).
\end{itemize}

The overlaps of various combinations of these
branes have been computed in \cite{us}, and in this paper, and
from them the open string spectrum on these
branes can be calculated.

The overlap of a 1-brane with a 1-brane was
calculated in \cite{us}, and the open string spectrum was found.
In the Lorentzian framework, the states on the
brane were given by the discrete states $\hat{D}_j^{+, \omega}$ and
the continuous states $\hat{C}_j^{\alpha, \omega}$. (This was originally found in \cite{ooguri}).
In the Euclidean theory, the open strings are in
the continuous representations of SL(2,C).

The open string spectrum between a 2-instanton and a 2-instanton
in the Euclidean theory is the same as that between a
1-brane and a 1-brane i.e. the strings are in
the continuous representations.
There is no Lorentzian analogue for the 2-instantons, since
these are not states in the Lorentzian theory.

The open string spectrum between a 2-instanton and a 1-brane
does not preserve the current algebra, and is the theory
of three free bosons, two with Neumann-Dirichlet conditions
and one with Neumann boundary conditions. 

The theory on the open strings connecting the three-brane to
itself is the theory
of three free bosons with Neumann boundary conditions.

The theory on the open strings connecting the three-brane to
a 1-brane is the theory
of three free bosons, two with Neumann boundary conditions
and the third with  Neumann-Dirichlet boundary conditions.
The theory on the open strings connecting the three-brane to
a 2-instanton is the same.

We have only found the spectrum of the various open strings. The
interactions of these strings are still largely unknown.
This is an important question that deserves further study.

\section{Acknowledgements}
This work was supported in part by  DOE
grant DE-FG02-96ER40559.
\section{Appendix}

$AdS_3$ is the universal cover of the group manifold $\SL$. A convenient
parametrization of a $\SL$ element is provided by the Euler angles:
\bea
g &=& e^{iu\sigma_2} e^{\rho \sigma_3} e^{iv \sigma_2} = \nonumber\\
  &=&
\left(
\begin{array}{lll}
\cos t\cosh\rho+\cos\phi\sinh\rho &\quad & \sin t\cosh\rho-
\sin\phi\sinh\rho\cr
-\sin t\cosh\rho-\sin\phi\sinh\rho &\quad &\cos t\cosh\rho-
\cos\phi\sinh\rho \cr
\end{array}
\right)
\eea
where $\sigma_i, i=1,2,3 $ are the Pauli matrices, and we define:
\beq
u = \frac{1}{2} (t+\phi)~~~~~~~~~~~~
v=  \frac{1}{2} (t-\phi)
\eeq
These are the cylindrical (global) coordinates of $AdS_3$. The metric is then:
\beq
ds^2 = - \cosh^2 \rho dt^2 + d\rho^2 + \sinh^2 \rho d\phi^2
\eeq

After a Wick rotation we obtain $\Eads$, which is the 
homogeneous space $SL(2,C)/SU(2)$.
We can parametrize $\Eads$ in the following
coordinate systems:

{\bf Cylindrical coordinates}. In
these coordinates
\bea
ds^2=d\rho^2 +\sinh^2\rho d\phi^2+\cosh^2\rho d\tau^2
\eea
The coordinates satisfy $0\le \rho<\infty, 0\le\phi<2\pi,
-\infty<\tau<\infty$.

\vskip 0.1cm

{\bf Poincare coordinates}. In
these coordinates
\bea
ds^2={1\over r^2}\left(dr^2+dx^2+d\tilde{\tau}^2\right)
\eea
The coordinates satisfy $0\le r<\infty, -\infty<x<\infty,
-\infty<\tilde{\tau}<\infty$.

{\bf 1-brane coordinates} or {\bf $AdS_2$ coordinates}
defined as
\bea
\sinh\psi=\sinh\rho\sinh\phi,\qquad
\cosh\psi\sinh\chi=-\sinh\rho\cos\phi,
\quad \tau=\tau
\eea
The metric in these coordinates is
\bea
ds^2=d\psi^2+\cosh^2\psi d\chi^2+\cosh^2\psi\cosh^2\chi d\t^2 
\eea

{\bf 2-instanton coordinates}, defined
as
\bea
\sinh\a=\cosh\rho\sinh\tau \quad\quad \cosh\a \cosh\b=\cosh\rho\cosh\tau,
\quad\phi=\phi
\eea
with $-\infty<\a<\infty, 0\le\b<\infty$.
The metric in these coordinates is
\bea
ds^2=d\a^2+\cosh^2\a d\b^2+\cosh^2\a \sinh^2 \b d\phi^2
\eea

{\bf 0-instanton coordinates}, defined
as
\bea
\cosh\l=\cosh\rho\cosh\tau\qquad \qquad \sinh\l\sin\s=\cosh\rho\sinh\tau,
\quad\phi=\phi
\eea
with $0\le\l<\infty, -{\pi\over 2}<\s\le{\pi\over 2}$.
The metric in these coordinates is
\bea
ds^2=d\l^2+\sinh^2\l d\s^2+\sinh^2\l\cos^2\s d\phi^2
\eea

\end{document}